\documentclass[11pt]{article}
\usepackage{a4,epsfig}
\newcommand{\xgo}{\mbox{$x_{\gamma}^{\rm obs}$}}
\renewcommand{\abstract}{}

\begin{document}
{

\begin{center}
\section*{Inclusive dijet photoproduction and the \\ resolved photon at THERA}

{\bf M. Wing}

\emph{McGill University, Montreal} 

{\tt wing@mail.desy.de}

\end{center}

\begin{center}
\subsubsection*{Abstract}
\end{center}

\abstract{A future $ep$ facility, THERA, where electrons of 250 GeV and protons of 920 GeV 
are collided could provide 
valuable information on the structure of the photon. With an increase in the centre-of-mass 
energy of a factor of 3 and an extension of the minimum photon energy fraction carried by 
the interacting parton of a factor of 10 compared to HERA, a new kinematic regime in the 
study of the photon will be opened. Inclusive dijet production has been studied and the 
potential gains the new collider would bring are discussed. The differences between current 
parametrisations of the photon structure in this new kinematic region are shown 
to be up to $50\%$. Comparisons of THERA's capabilities are made with what HERA can 
currently produce and how it complements $e^+e^-$ colliders addressed.}

\bigskip

\section{Introduction}

Using the proton accelerator ring from HERA and the lepton accelerator chain for the proposed 
linear collider, TESLA, a centre-of-mass energy for $ep$ collisions of $\sim$ 1 TeV could be 
achieved~\cite{TDR}. The proposed facility, THERA, using protons of 920 GeV and leptons of 
250 GeV would, due to the increase in lepton beam energy, greatly extend the kinematic regime 
currently accessible to HERA~\cite{TDR,kinematics}.

The measurement of jet photoproduction in $ep$ collisions allows the structure 
of the photon, emitted from the incoming lepton, to be probed. In Leading Order (LO) quantum 
chromodynamics (QCD), two types of processes contribute to jet photoproduction: the direct 
photon 
process, in which the photon itself interacts with a parton from the proton and the resolved 
photon process in which the photon acts as a source of partons, one of which interacts with 
a parton from the proton. Examples of these processes are shown in Fig.~\ref{feynman}, where 
Fig.~\ref{feynman}a shows the boson-gluon fusion direct process and Fig.~\ref{feynman}b 
gluon-gluon fusion resolved process. As can be seen from Fig.~\ref{feynman}, photoproduction 
processes depend on both the structure of the photon and proton; the  
cross-section, $d\sigma_{\gamma p \rightarrow cd}$ for the production of two partons being,
\begin{equation}
d\sigma_{\gamma p \rightarrow cd} = \sum_{ab} \int_{x_p} dx_p
\int_{x_\gamma} dx_\gamma f_{p \rightarrow b}(x_p, \mu^2) f_{\gamma
\rightarrow a}(x_\gamma, \mu^2) d\hat{\sigma}\mathit{_{ab \rightarrow cd}},
\label{general_xsec}
\end{equation}
where $f_{p \rightarrow b}(x_p, \mu^2)$ describes the proton parton density of a parton 
of momentum 
fraction, $x_p$, $f_{\gamma \rightarrow a}(x_\gamma, \mu^2)$ the photon parton density of 
a parton 
of momentum fraction, $x_\gamma$, both at some renormalisation and factorisation scale, 
$\mu^2$, and $d\hat{\sigma}\mathit{_{ab \rightarrow cd}}$ is the perturbatively calculable 
short distance cross-section. As next-to-leading (NLO) order programs reasonably describe 
jet production, 
one can choose a region of phase space where the proton structure is well constrained and 
the only uncertainty, as can be seen in Eq.~\ref{general_xsec}, is then the photon structure 
function. It is noted (although not studied further) that dijet production at HERA and THERA 
could also provide information on the proton structure using high transverse energy or 
(very) forward jets. This would provide information on the parton densities at high values 
of $x_p$ at large scales, as in $p\bar{p}$ collisions, which are not well constrained by 
the $F_2^p$ DIS measurements.

\begin{figure}[htp]
\unitlength=1mm
\begin{picture}(0,0)(100,100)
\put(143,30){\bf \Large{(a)}}
\put(206,30){\bf \Large{(b)}}
\end{picture}
\begin{center}
~\epsfig{file=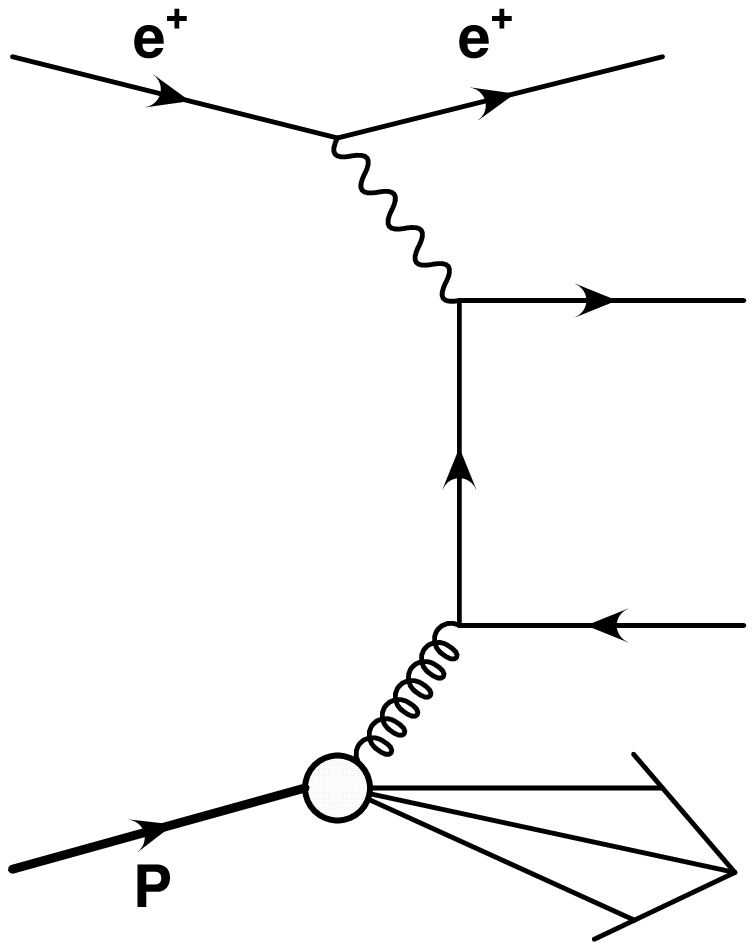,height=6.5cm}
\hspace{1cm}~\epsfig{file=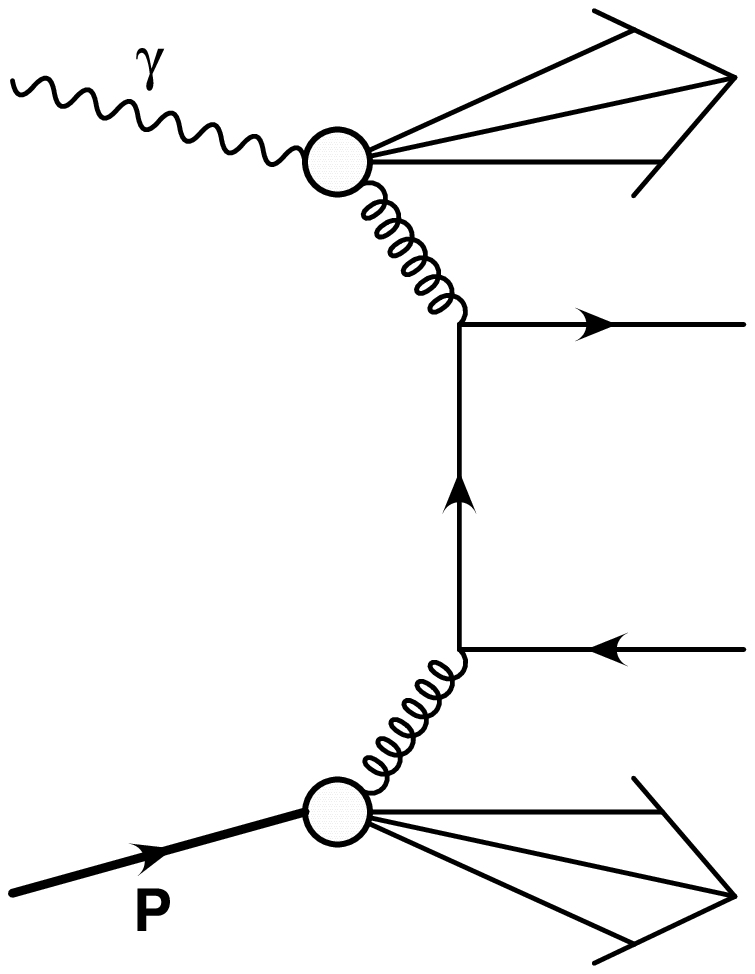,height=6.5cm}
\end{center}
\caption{Examples of LO (a) direct photon and (b) resolved photon processes in $ep$ 
collisions.}
\label{feynman}
\end{figure}

Measurements of the photon structure function, $F_2^\gamma$, at $e^+e^-$ experiments have 
constrained the quark density in the region $10^{-3}<x_\gamma<0.5$~\cite{lep_f2}. 
Photoproduction can provide further constraints on the quark density at high $x_\gamma$ and 
on the gluon density  at all $x_\gamma$. In jet photoproduction, unlike in $e^+e^-$ 
scattering, there is direct access to the gluon density in the photon at LO. The poorly 
constrained gluon density is expected to become more significant with decreasing $x_\gamma$. 
Measurements have been made at HERA, which probe these uncertain parts of the structure of the 
photon~\cite{zeus_lowet,zeus_highet,zeus_prelim,h1_gluon}. An electron beam energy of a 
factor of 10 higher for THERA than for HERA gives an extension in the minimum $x_\gamma$ 
also by a factor of 10~\cite{kinematics}. THERA would lead to an extension of the kinematic 
range into a region where accurate measurements of the gluon density could be made and the 
rise in the structure function, $F_2^\gamma$, thoroughly tested. 

\section{Photoproduction from HERA to THERA}

Currently at HERA, the highly asymmetric beam energies provide problems in measuring 
resolved photon processes. In general, there is a strong tendency for events to be 
very forward in the proton beam direction. This means that many events fall outside the 
acceptance of the current detectors or are mixed up with the proton remnant making their 
measurement difficult. This is particularly acute in the case of resolved photon 
processes in which jets are generally produced in the forward direction; direct photon 
processes generally being central. The reduced asymmetry THERA would provide will 
significantly change the topology of photoproduction events. Direct 
photon events would then be concentrated 
in the rear direction and resolved photon processes in the central and forward parts of 
the detector. This would then increase the acceptance for resolved photon events and 
hence improve the measurements.

Measurements of dijet photoproduction at HERA can be (somewhat artificially) categorised 
into those with low transverse energy jets, low$-E_T^{\rm jet}$, and those with 
high$-E_T^{\rm jet}$ 
which start at around 5 GeV and 10 GeV, respectively. The advantage of the low$-E_T^{\rm jet}$ 
measurements is the increased resolved photon cross section, in particular at low$-x_\gamma$. 
The measurements suffer, however, from a lack of understanding of soft underlying physics 
which either leads to results with large systematic uncertainties or inconclusive comparisons 
with theory. At high$-E_T^{\rm jet}$, the effect of soft underlying physics is greatly reduced, 
but so is the resolved cross section at low$-x_\gamma$. This can be seen by considering the 
observable \xgo, which is the fraction of the photon's energy producing the two jets of 
highest transverse energy~\cite{xgo};
\begin{equation}
\xgo = \frac{E_T^{\rm jet1}{\rm e}^{- \eta ^{\rm jet1}} 
           + E_T^{\rm jet2}{\rm e}^{- \eta ^{\rm jet2}}}{2yE_e},
\label{eq:xgamma}
\end{equation}
where $\eta^{\rm jet}$ is the pseudorapidity of the jet and $E_e$ is the electron energy and 
$y$ the fraction of the electron's energy carried by the photon. 

Recent results from HERA are shown for low$-E_T^{\rm jet}$ events ($E_T^{\rm jet}>6$ GeV) in 
Fig.~\ref{xgamma_hera}a and for high$-E_T^{\rm jet}$ events ($E_T^{\rm jet}>14$ GeV) in 
Fig.~\ref{xgamma_hera}b. Shown are the detector-level distributions for \xgo \ compared 
to Monte Carlo (MC) models. In Fig.~\ref{xgamma_hera}a, the data is compared to MC models 
with and without multiparton interactions (MPI). Here it can be seen that the MC without 
MPI greatly underestimates the data at low$-\xgo$, and that a significantly increased 
cross-section is observed in the MC with the inclusion of MPI. The data is, however, still 
poorly described by the MC at this low$-\xgo$ region for jets of low$-E_T^{\rm jet}$. In 
Fig.~\ref{xgamma_hera}b, the data compares well, over the full \xgo \ range, with MC models 
containing no MPI, due to the increased $E_T^{\rm jet}$. The significant decrease in the 
number of low$-\xgo$ events at high$-E_T^{\rm jet}$ is also apparent.

\begin{figure}[htp]
\unitlength=1mm
\begin{picture}(0,0)(100,100)
\put(170,75){\bf \Large{(a)}}
\put(235,75){\bf \Large{(b)}}
\end{picture}
\begin{center}
~\epsfig{file=xg-bob.epsi,height=7cm}
~\epsfig{file=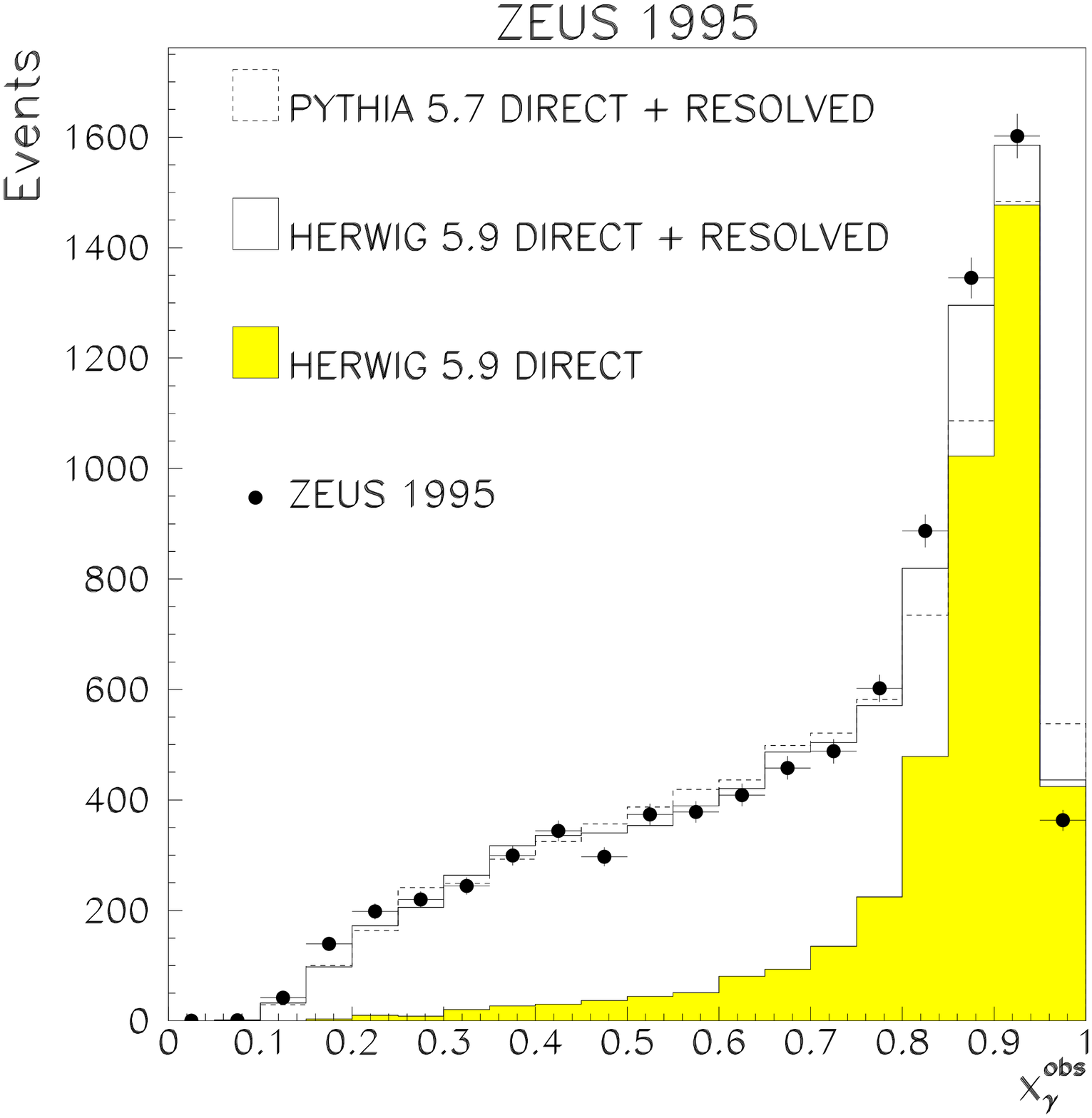,height=7cm}
\end{center}
\vspace{-0.5cm}
\caption{Detector-level distributions of \xgo \ at HERA for (a) $E_T^{\rm jet}>6$ GeV 
and (b) $E_T^{\rm jet}>14$~GeV. In (a), the data points are compared to {\sc Herwig} 
MC~\cite{herwig} with (solid line) and without (dotted line) multiparton interactions and 
{\sc Pythia} MC~\cite{pythia} predictions also with multiparton interactions (dashed line). 
In (b), the 
data points are compared to {\sc Herwig} MC (solid line) and {\sc Pythia} MC (dashed line) 
predictions, both without multiparton interactions. In both (a) and (b) the shaded histogram 
indicates the contribution from direct photon processes in {\sc Herwig}.}
\label{xgamma_hera}
\end{figure}

With the increased lepton beam energy THERA could provide the optimum scenario in which 
measurements could be made at relatively high$-E_T^{\rm jet}$ and low$-\xgo$.

\section{Cross section definition}

A realistic approach has been performed to evaluate the extent to which dijet production 
at THERA can yield new information on the structure of the photon. Therefore a kinematic 
range has been considered which could possibly be measured rather than choosing the full 
phase space. The kinematic region is based on a high$-E_T^{\rm jet}$ 
measurement from the ZEUS collaboration~\cite{zeus_highet,zeus_prelim}. 

To define the region to be photoproduction, a value of the photon virtuality, $Q^2$, of 
less than 1 ${\rm GeV}^2$ has been applied. Currently at HERA, the requirement that the 
scattered lepton is not seen in the calorimeter implicitly leads to the condition on the 
photon virtuality. To achieve the same $Q^2$ requirement with the increased lepton beam 
energy, the angle of scatter will be much smaller. This would mean that the calorimeter 
(or some other detector) would need to be positioned very close (in fact to within 
$\sim0.5^\circ$) to the beam-pipe in the rear direction to achieve the same $Q^2$ requirement 
using the anti-tag condition. The inelasticity, $y$, 
is chosen to be within the region, $0.2<y<0.85$, which corresponds to a photon-proton 
centre-of-mass energy of between 429 GeV and 884 GeV. The jets are required to have 
transverse energies, $E_T^{\rm jet1,2}>14,11$ GeV in the region of pseudorapidity, 
$-1<\eta^{\rm jet}<2$. Increasing the cut on the transverse energy was also considered. 
Allowing the jets to go more forward in pseudorapidity would increase the resolved photon 
cross section and would hopefully be possible depending on the detector configuration. At 
HERA where the proton beam energy is very much greater than the lepton beam energy, there 
are very few jets with a pseudorapidity less than -1. At THERA the distribution of jets 
would be more symmetric in pseudorapidity and relaxing this cut to, say -2 or -2.5 would 
also be an option. For this analysis, the value of -1 is retained to try and enhance the 
(more forward going) resolved photon component in a part of phase space which could 
definitely be measured. Extension and modification of these requirements could be 
considered in future studies.

For studying the potential of $ep$ collisions at $\sim 1$ TeV, the cross sections were 
produced using NLO code from Frixione and Ridolfi~\cite{frix_et_rid}. As the proton 
structure function is well constrained in the region of $x_p$ under study (approximately 
$10^{-2}<x_p<10^{-1}$), only the CTEQ-4M~\cite{cteq} parton distribution function was 
considered. The renormalisation and factorisation scales, $\mu$, were set to be equal to 
$E_T^{\rm jet1}$. To test the sensitivity of the cross sections, three photon parton  
density parametrisations were used; GS96-HO~\cite{gs96}, GRV-HO~\cite{grvho} and 
AFG-HO~\cite{afgho}. 
Jets are defined using the longitudinally invariant $k_T$-clustering 
algorithm~\cite{cluster} in the inclusive mode~\cite{inclusive_mode}.

\section{Results}

The kinematic requirements for the analysis performed at HERA restrict the minimum \xgo \ 
to be $\sim0.07$, where both jets are of minimum transverse energy and have a pseudorapidity 
of 2. At THERA the minimum \xgo \ is reduced by the same factor as the lepton beam energy 
is increased to $\sim0.008$. 
Figure~\ref{xgamma}a shows the cross-sections in \xgo \ for HERA and THERA for the kinematic 
region stated. At HERA, the events are concentrated at high$-$\xgo \ arising predominantly 
from direct photon events. At THERA, the events are concentrated at low$-$\xgo \ characteristic 
of resolved photon events, with a very small cross-section at \xgo$>$0.75 which is 
taken to be a direct photon enriched region. Having seen that THERA greatly improves the 
potential for studying the resolved photon, it is then interesting to see if the new 
kinematic region shows sensitivity to the current parametrisations of the structure of the 
photon. Results for the three different photon parton density parametrisations are shown in 
Fig.~\ref{xgamma}a and their relative difference in Fig.~\ref{xgamma}b. It can be seen 
that the prediction using the GS96-HO parametrisation gives the highest and using the 
AFG-HO parametrisation the lowest cross section. The prediction with GS96-HO is up 
to $35\%$ higher than that of GRV-HO and $50\%$ higher than that of AFG-HO at low$-$\xgo. 
The difference between the results based on the three parton density parametrisations 
decreases with increasing \xgo. It 
should be noted that at the lowest values of \xgo \ shown here, there exist no measurements 
from LEP at the scale considered ($\sim 200~{\rm GeV^2}$). 

\begin{figure}[htp]
\unitlength=1mm
\begin{picture}(0,0)(100,100)
\put(225,75){\bf \Large{(a)}}
\put(225,-27){\bf \Large{(b)}}
\end{picture}
\begin{center}
~\epsfig{file=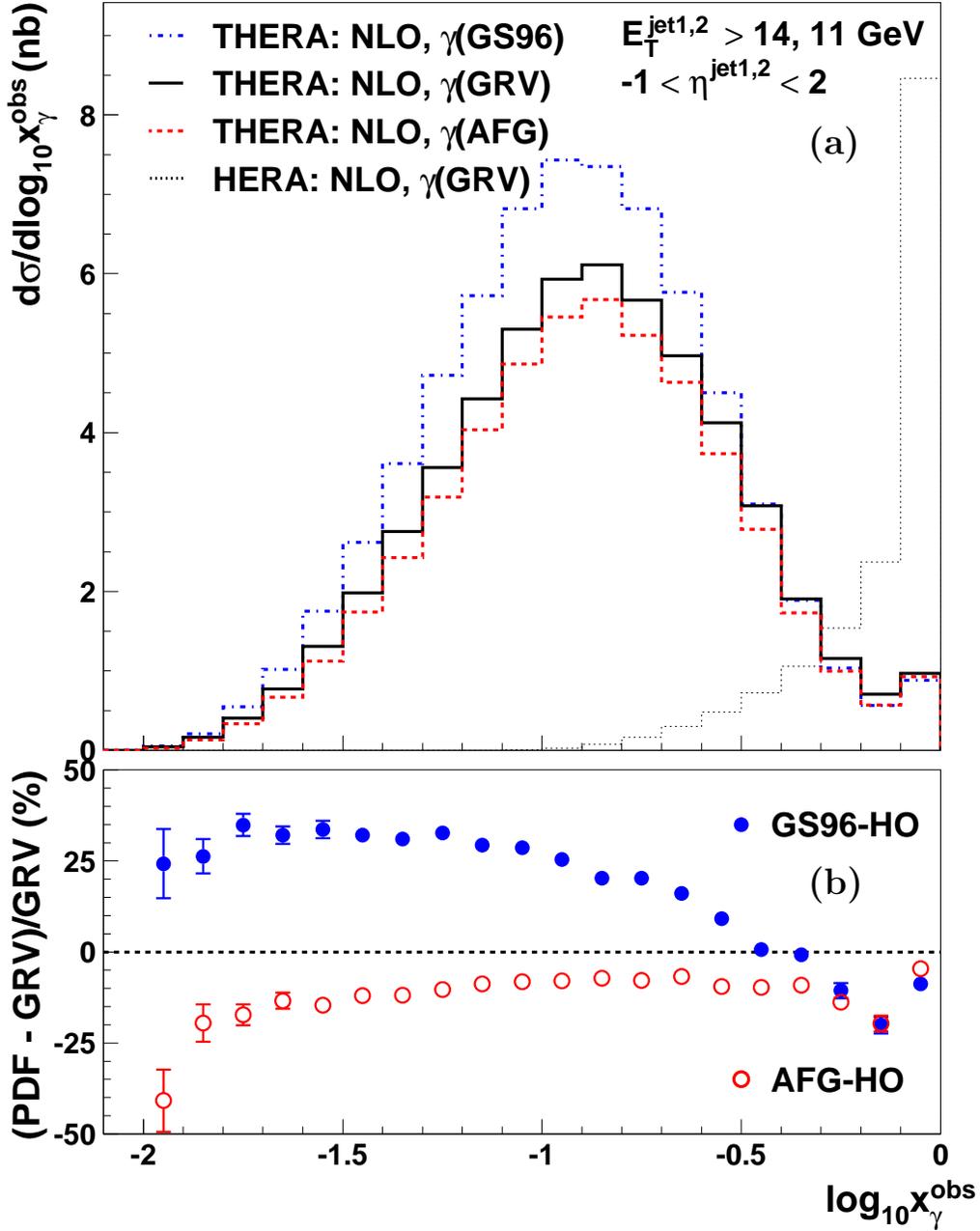,height=17.cm}
\end{center}
\vspace{-0.5cm}
\caption{(a) The differential cross section, $d\sigma/d{\rm log_{10}}\xgo$ for inclusive 
dijet photoproduction at HERA and THERA as predicted by a NLO calculation. For the kinematic 
range, $Q^2<1~{\rm GeV}^2$, $0.2<y<0.85$ with two jets, $E_T^{\rm jet1,2}>14,11$~GeV and 
$-1<\eta^{\rm jet}<2$, the prediction for HERA using the GRV-HO photon parametrisation is 
shown as the dotted line. For THERA with the same kinematic cuts, three photon parton 
density parametrisations are shown; GS96-HO (dot-dashed line), GRV-HO (solid line) and 
AFG-HO (dashed line).
In (b) the percentage differences in the cross-sections between the three predictions for 
THERA are shown as a function of ${\rm log_{10}}\xgo$. The relative difference of the 
predictions using GS96-HO (solid points) and AFG-HO (open points) with respect to GRV-HO is 
displayed.}
\label{xgamma}
\end{figure}

As can be seen from Eq.~\ref{eq:xgamma}, the minimum \xgo \ increases with increasing 
$E_T^{\rm jet}$. The relative fraction of low to high$-$\xgo \ and hence resolved photon 
to direct photon processes also decreases with increasing $E_T^{\rm jet}$. Nevertheless, 
differences between the predictions with GS96-HO and AFG-HO are up still to $30\%$ at a 
minimum cut-off in the transverse energy of the leading jet of 29~GeV.

\bigskip

Cross-sections as a function of the transverse energy of the leading jet, $E_T^{\rm jet1}$,  
in different regions of pseudorapidity of the jets are shown in Fig.~\ref{et_thera}. It 
is shown that the differences between the three predictions are concentrated at 
low$-E_T^{\rm jet1}$ and forward pseudorapidity. At more central and rear values of 
pseudorapidity and higher transverse energies, the predictions converge.

\begin{figure}[htp]
\unitlength=1mm
\begin{picture}(0,0)(100,100)
\put(222,87){\bf \Large{(a)}}
\put(222,-10){\bf \Large{(b)}}
\end{picture}
\begin{center}
~\epsfig{file=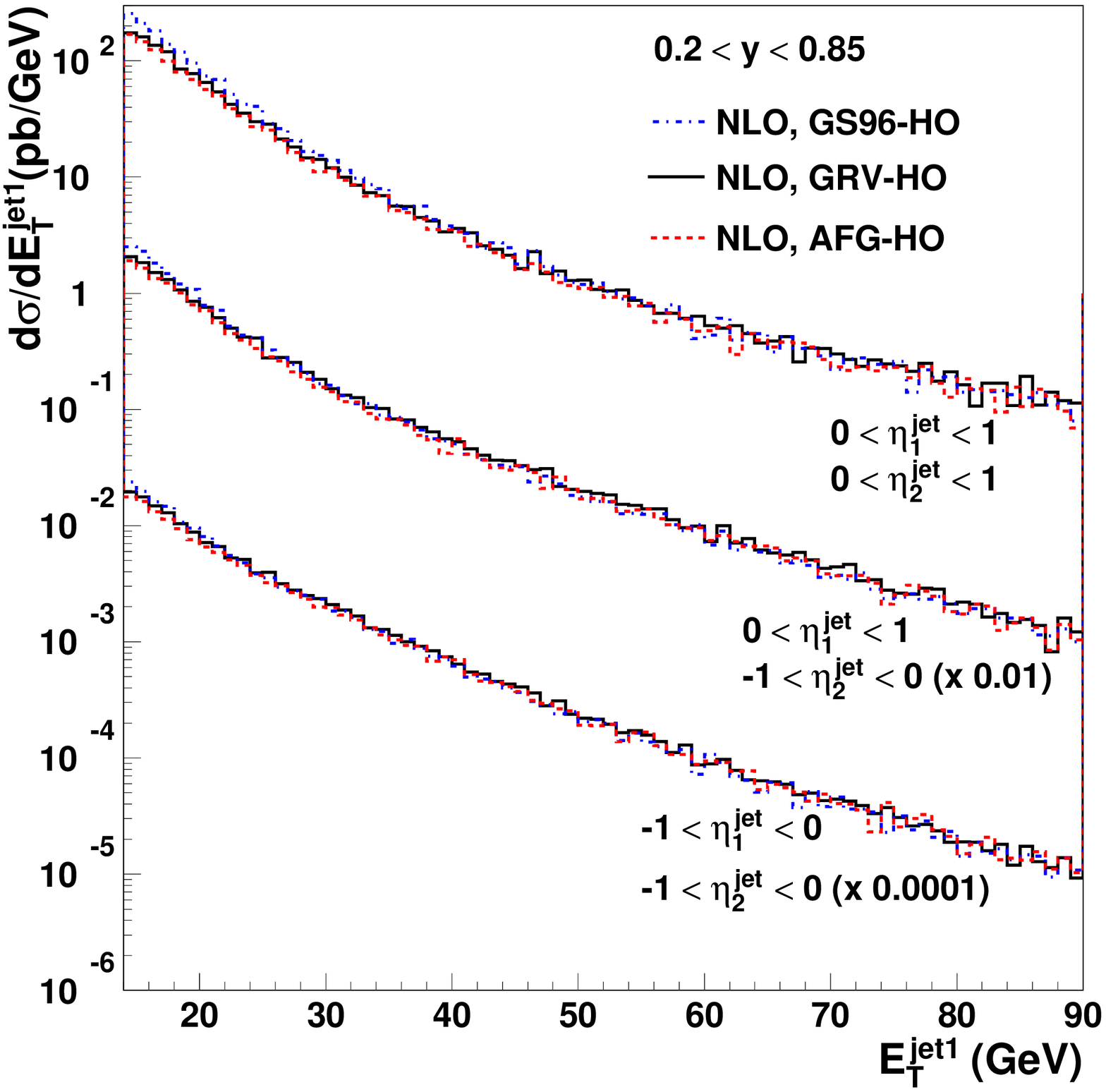,height=9.7cm,width=11cm}
~\epsfig{file=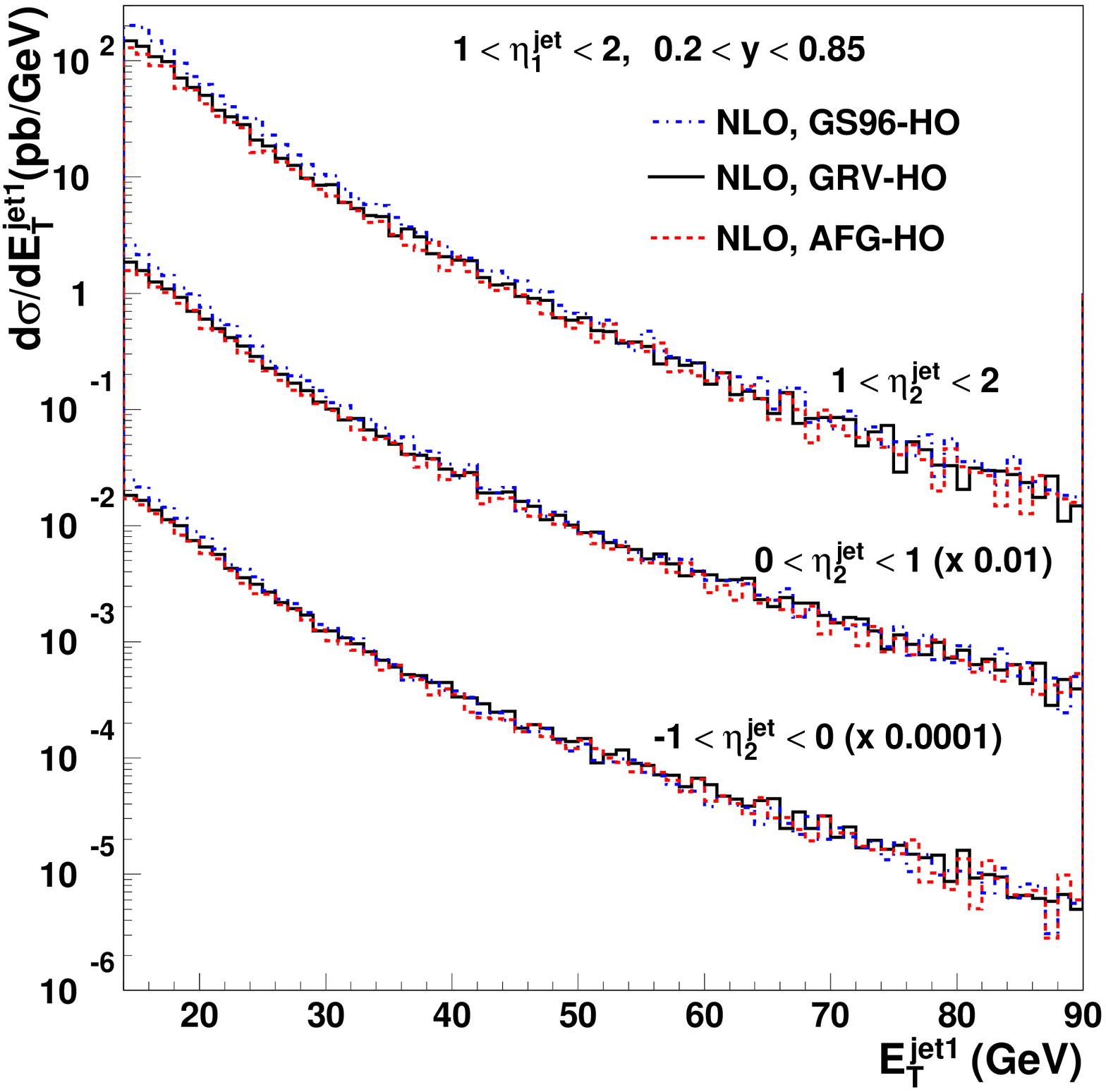,height=9.7cm,width=11cm}
\end{center}
\vspace{-0.5cm}
\caption{(a) The differential cross section, $d\sigma/dE_T^{\rm jet1}$ for inclusive dijet 
photoproduction at THERA as predicted by a NLO calculation. For the kinematic range as in 
Fig.~\ref{xgamma}, the prediction for THERA using three photon parton density parametrisations 
are shown; GS96-HO (dot-dashed line), GRV-HO (solid line) and AFG-HO (dashed line). 
The three different sets of curves represent jets in different regions of pseudorapidity.
(b) The differential cross section, $d\sigma/dE_T^{\rm jet1}$ for inclusive dijet 
photoproduction at THERA as predicted by a NLO calculation when one jet is in the region 
$1<\eta^{\rm jet1}<2$ and the other jet is in three different regions of pseudorapidity.}
\label{et_thera}
\end{figure}

\bigskip

The cross-section as a function of the pseudorapidity of the second jet, $\eta_2^{\rm jet}$, 
is expected to be sensitive to the structure of the photon~\cite{zeus_highet}. In 
Fig.~\ref{eta}, this cross section is shown in three regions of pseudorapidity of the 
first jet, $\eta_1^{\rm jet}$. Again, it can be seen that the predictions differ most 
significantly, up to $50\%$, when both jets are forward. It can also be seen that in the 
region where both jets are forward the direct photon enriched region, $\xgo>0.75$, 
is negligible. As for the cross section in \xgo, the predictions for the GRV-HO and 
AFG-HO parametrisation are very similar in shape and differ in magnitude by roughly 
$10\%$. At higher transverse energies, $E_T^{\rm jet1}>$ 29 GeV, a difference of $30\%$ 
persists when both jets are forward.

\bigskip

To assess the significance of the differences observed between the predictions with the 
three different parametrisations, the renormalisation and factorisation scale 
uncertainties were evaluated. The cross sections using the GRV-HO parametrisation were 
calculated with the scale doubled and halved. For the cross-sections as a function of 
$\eta_2^{\rm jet}$, the relative differences to the central values are shown. The differences 
are of the order of $10-15\%$, which is small compared with the differences between the 
parametrisations. A difference of $15\%$ was also observed for the cross-section as a 
function of \xgo. 

There exist many other calculations of jet photoproduction at HERA~\cite{other_NLO} all 
of which (including the one used in this article) have been shown to agree within 
$5-10\%$~\cite{zeus_highet,nlo_compare}. From this and the estimation of the scale 
uncertainty, it can be seen that the differences in the photon parton denisty parametrisations 
are much larger than other uncertainties.

\begin{figure}[htp]
\unitlength=1mm
\begin{picture}(0,0)(100,100)
\put(240,85){\bf \Large{(a)}}
\put(240,-10){\bf \Large{(b)}}
\end{picture}
\begin{center}
~\epsfig{file=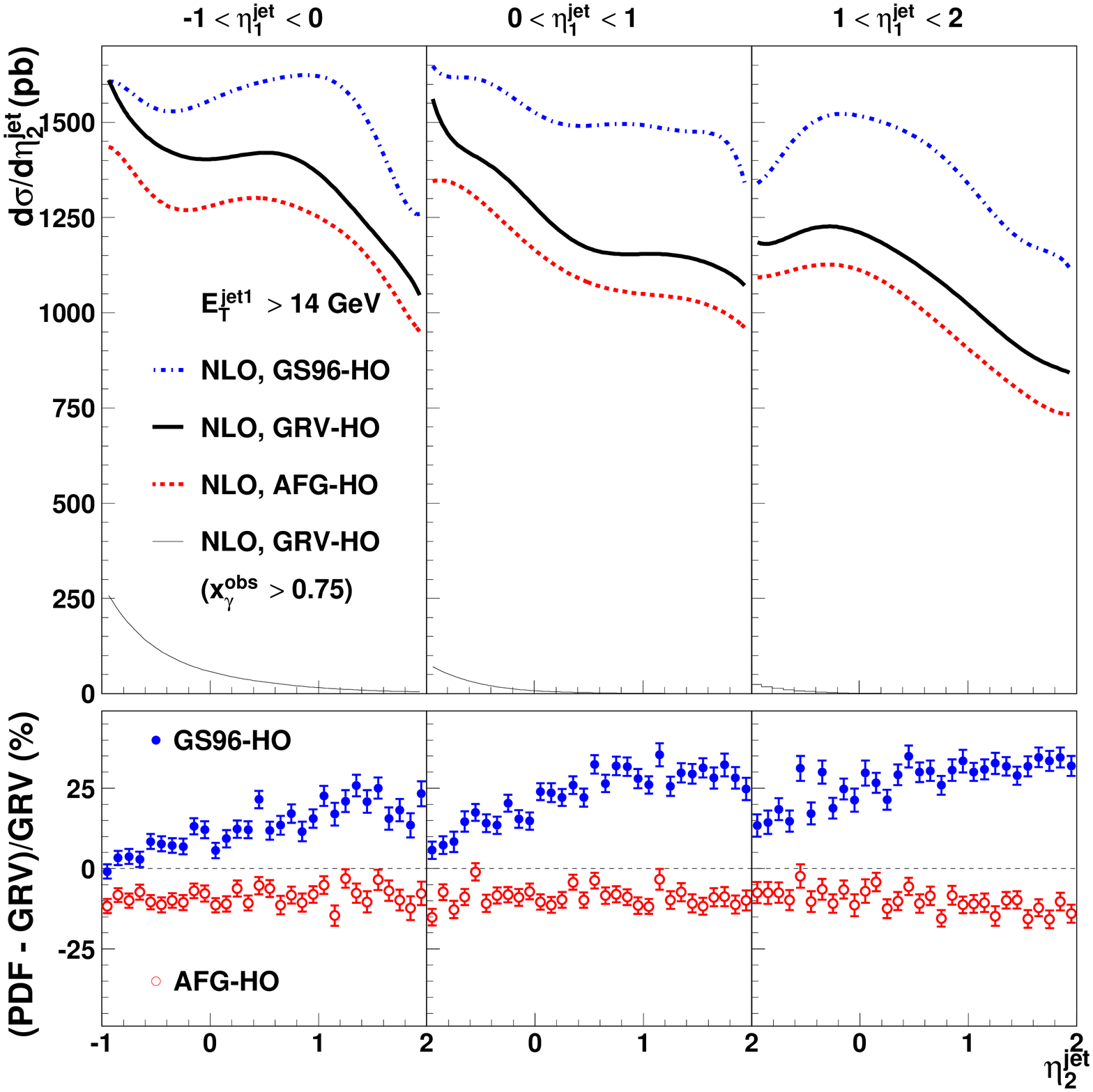,height=14.cm}
\end{center}
\vspace{-0.5cm}
\caption{(a) The differential cross section, $d\sigma/d\eta_2^{\rm jet}$ for inclusive 
dijet photoproduction at THERA as predicted by a NLO calculation. For the kinematic range, 
$Q^2<1~{\rm GeV}^2$, $0.2<y<0.85$ with two jets, $E_T^{\rm jet1,2}>14,11$~GeV and 
$-1<\eta^{\rm jet}<2$, the prediction for THERA using three photon parton density 
parametrisations are 
shown; GS96-HO (dot-dashed line), GRV-HO (solid line) and AFG-HO (dashed line). One jet 
is restricted to be in a rear, central or forward direction. The prediction for $\xgo>0.75$ 
is also shown as the thin solid line. In (b) the percentage differences in the 
cross-sections between the three predictions for THERA are shown as a function of 
$\eta_2^{\rm jet}$. The relative difference of the predictions using GS96-HO (solid points) 
and AFG-HO (open points) with respect to GRV-HO is displayed.}
\label{eta}
\end{figure}

\begin{figure}[htp]
\begin{center}
~\epsfig{file=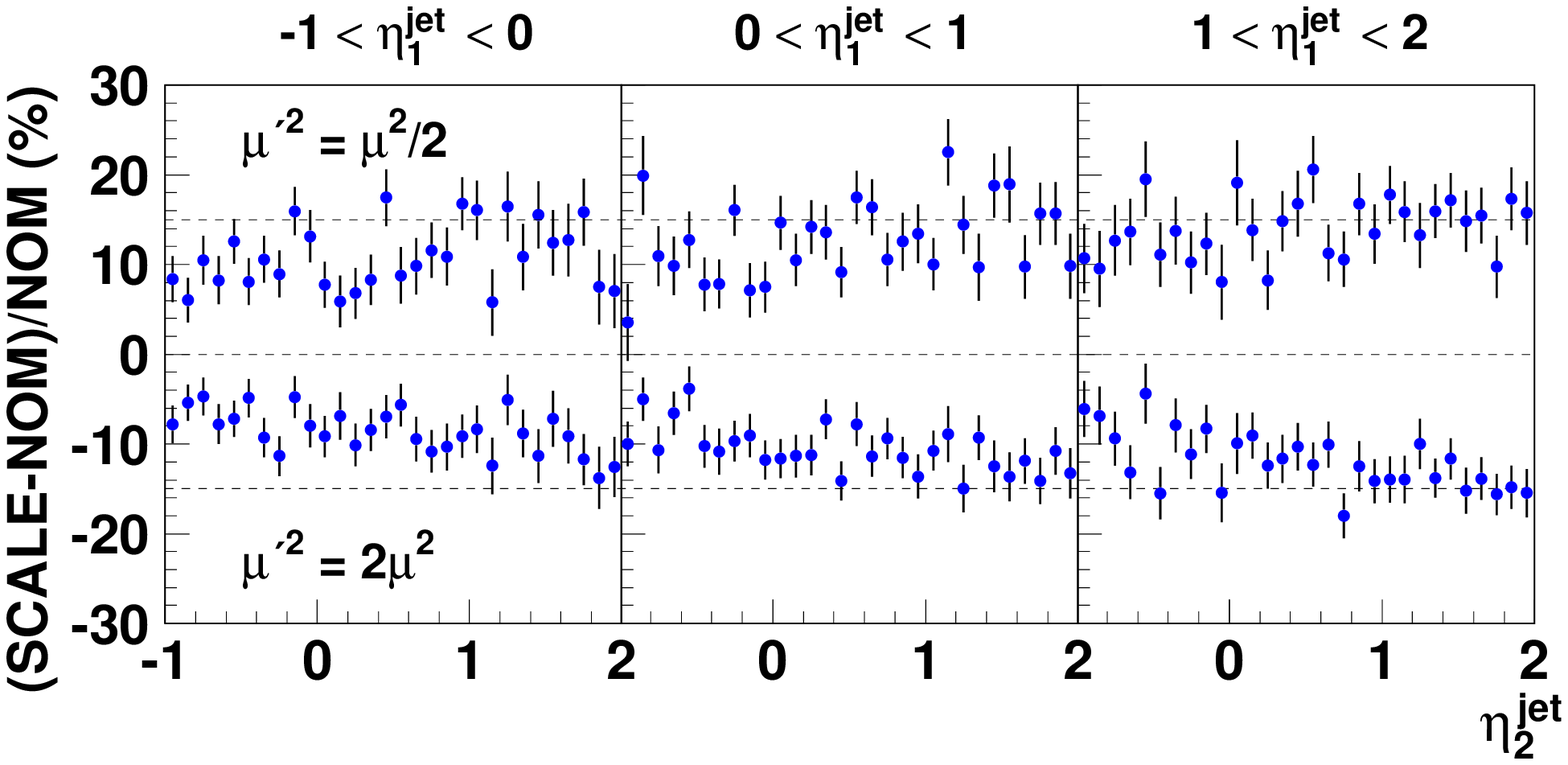,height=6.5cm}
\end{center}
\vspace{-0.5cm}
\caption{The percentage differences in the cross-sections from Fig.~\ref{eta} when varying 
the renormalisation and factorisation scales by a factor of 0.5 and 2. The photon parton density 
parametrisation used was GRV-HO. The dashed line indicates $\pm 15\%$.}
\label{eta_scale}
\end{figure}

\section{Discussion}

The potential of THERA with respect to what can be done at HERA and at LEP in the field 
of the photon structure is as follows. For a given transverse energy, it can achieve 
lower values of $x_\gamma$ than is accessible at either of the current facilities. THERA 
would have the potential to measure the structure of the photon in the region where 
the structure function is predicted to rise with decreasing $x_\gamma$. Indications for 
the rise have be seen at LEP and HERA although the measurements have large errors. THERA 
would, however, be able to make more precise measurements by being able to achieve the 
factor of 10 smaller in the minimum \xgo \ for a given $E_T^{\rm jet}$ compared with HERA. 
It has also been shown elsewhere that the accessible maximum average transverse energy of a dijet 
system is 2-3 times larger 
at THERA than at HERA with roughly 150 GeV and 225 GeV being reachable at the $ep$ and 
$\gamma p$ options, respectively~\cite{mklasen}.

Forward detectors would also be desirable for measurements of the proton structure function 
at high $x_p$. For studying values of $x_p>0.5$, for example, jets of $E_T^{\rm jet}$ greater 
than roughly 20~GeV would need to be detected at pseudorapidity values, $\eta^{\rm jet}=3$. 
Jets of this energy would be produced copiously; considering those more central would 
require much larger energy.

Heavy quarks in dijet production have also been studied as a tool for constraining the 
structure of the photon and are discussed in detail elsewhere~\cite{hq_php_thera}.

With respect to the linear 
collider, TESLA, lower values of $x_\gamma$ can also be reached at THERA but only at very forward 
values of pseudorapidity~\cite{kinematics}. Development of a detector for THERA which can 
measure jets in the forward direction would therefore be very desirable. THERA would also 
complement a linear collider in the measurement of the structure of the photon in much the 
same way as HERA complements LEP; for example in directly constraining the gluon density and 
testing the universality of the parton distribution functions.

\section{Summary}

It has been shown that cross sections in the kinematic region open to THERA with a 
centre-of-mass energy of $\sim$ 1 TeV have a large sensitivity to the structure of the 
photon. Three currently available parametrisations lead to cross sections which vary by up 
to $50\%$ in certain regions of the phase space considered. THERA would provide a new 
kinematic region in the measurement of the structure of the photon and complement 
measurements from the proposed linear collider.

\section*{Acknowledgements}

I would like to thank S. Frixione and G. Ridolfi for making their NLO code freely available 
on the web and in particular S. Frixione who answered many questions on how to use the code 
such that I could use it for this study.

}

\begin{thebibliography}{99}

\bibitem{TDR}
THERA study group, Electron-proton scattering at $\sqrt{s}\sim1$ TeV. Physics and 
experimentation with THERA, TESLA technical design report, DESY (2001) Appendix.

\bibitem{kinematics}
M. Krawczyk, S. S\"{o}ldner-Rembold, M. Wing, ``Kinematics of Photoproduction'', 
The THERA Book, DESY-LC-REV-2001-062 (2001).

\bibitem{xgo}
ZEUS Collaboration, M. Derrick et al., {\em Phys. Lett.} {\bf B 348} 665 (1995)

\bibitem{herwig}
G. Marchesini et al., {\em Comp. Phys. Commun.} {\bf 67} (1992) 465.

\bibitem{pythia}
H.-U. Bengtsson and T.~Sj{\"o}strand, {\em Comp. Phys. Commun.} {\bf 46} (1987) 43.

\bibitem{lep_f2}
S. S\"{o}ldner-Rembold, ``Experimentalists Summary'', talk given at Photon 2000, Ambleside, 
UK, 26-31 August 2000.

\bibitem{zeus_lowet}
ZEUS Collaboration, J. Breitweg et al., {\em Euro. Phys. J.} {\bf C 11} (1999) 35.

\bibitem{zeus_highet}
ZEUS Collaboration, J. Breitweg et al., {\em Euro. Phys. J.} {\bf C 11} (1998) 109.
 
\bibitem{zeus_prelim}
ZEUS Collaboration, ICHEP$-$418, Submitted to ICHEP2000 , Osaka, Japan.\\
{\tt http://www-zeus.desy.de/$\sim$schlenst/conf/osaka\_paper/QCD/dijetpho.ps.gz};\\
ZEUS Collaboration, ${\rm EPS-540}$, Submitted to the EPS High Energy Physics 99 conference, 
Tampere, Finland.\\ {\tt http://www-zeus.desy.de/eps99/eps99\_540.ps.gz}
 
\bibitem{h1_gluon}
H1 Collaboration, C. Adloff et al., {\em Phys. Lett.} {\bf B 483} (2000) 36.

\bibitem{frix_et_rid}
S. Frixione, Z. Kunszt and A. Signer, {\em Nucl. Phys.}~{\bf B467}~(1996)~399;\\ 
S. Frixione, {\em Nucl. Phys.} {\bf B507} (1997) 295; \\ 
S. Frixione and G. Ridolfi, {\em Nucl. Phys.} {\bf B507} (1997) 315.

\bibitem{cteq}
H. L. Lai et al., {\em Phys. Rev.} {\bf D 55} (1997) 1280.

\bibitem{gs96}
L. E. Gordon and J. K. Storrow, {\em Nucl. Phys} {\bf B 489} (1997) 405.

\bibitem{grvho}
M. Gl\"{u}ck, E. Reya and A. Vogt, {\em Phys. Rev.} {\bf D 45} (1992) 3986;\\
M. Gl\"{u}ck, E. Reya and A. Vogt, {\em Phys. Rev.} {\bf D 46} (1992) 1973.

\bibitem{afgho}
P. Aurenche, J. Guillet and M. Fontannaz, {\em Z. Phys.} {\bf C 64} (1994) 621.

\bibitem{cluster}
S. Catani et al., {\em Nucl. Phys.} {\bf B 406} (1993) 187.

\bibitem{inclusive_mode}
S. D. Elllis and D. E. Soper, {\em Phys. Rev.} {\bf D 48} (1993) 3160.

\bibitem{other_NLO}
B. W. Harris and J. F. Owens, {\em Phys. Rev.} {\bf D 56} (1997) 4007;\\
B. W. Harris and J. F. Owens, {\em Phys. Rev.} {\bf D 57} (1998) 5555;\\
M. Klasen, and G. Kramer, {\em Z. Phys.} {\bf C 76} (1997) 67;\\
M. Klasen, T. Kleinwort and G. Kramer, {\em Euro. Phys. J. direct} {\bf C 1} (1998) 1;\\
P. Aurenche et al., Proceeding of the Workshop ``Future Physics at HERA'', (1996) 570;\\
P. Aurenche et al., {\em Euro Phys. J.} {\bf C 17} (2000) 413;\\
G. Kramer and B. P\"{o}tter, {\em Euro. Phys. J.} {\bf C 5} (1998) 665;\\
B. P\"{o}tter, {\em Euro. Phys. J.} {\bf C 5} (1999) 1;\\
B. P\"{o}tter, {\em Comp. Phys. Commun.} {\bf 119} (1999) 4.

\bibitem{nlo_compare}
B. W. Harris, M. Klasen and J. Vossebeld, Proceedings 
of the Workshop on ``Monte Carlo Generators for HERA Physics'', Hamburg, Germany (1998) 
171, {\tt  hep-ph/9905348};\\
B. P\"{o}tter, {\em Comp. Phys. Commun.} {\bf 133} (2000) 105.

\bibitem{mklasen}
M. Klasen, ``Jet photoproduction at THERA'', The THERA Book, DESY-LC-REV-2001-062 (2001) 
{\tt  hep-ph/0103091}.

\bibitem{hq_php_thera}
P. Jankowski, M. Krawczyk and M. Wing, ``Heavy quarks in photoproduction at THERA'', 
The THERA Book, DESY-LC-REV-2001-062 (2001).

\end{thebibliography}
\end{document}